\documentstyle[12pt,psfig,epsfig]{article}

\newcommand {\be}{\begin{equation}}
\newcommand {\ee}{\end{equation}}
\newcommand {\bea}{\begin{eqnarray}}
\newcommand {\eea}{\end{eqnarray}}
\newcommand {\bx}{{\bf x}}
\newcommand {\by}{{\bf y}}
\newcommand {\bX}{{\bf X}}
\newcommand {\bY}{{\bf Y}}

\begin{document}

\begin{titlepage}

\title{Learning Driver-Response Relationships from
  Synchronization Patterns}

\vspace{2cm}
\author{R. Quian Quiroga$^\dagger$, J. Arnhold, and P. Grassberger \\
{\sl John von Neumann Institute for Computing,}\\
{\sl Forschungszentrum J\"ulich GmbH,}\\
{\sl D - 52425 J\"ulich, Germany}}

\maketitle
\vspace{3cm}
\hspace{0.3cm}   05.45.Xt

\vspace{0.5cm}
$^\dagger$ corresponding author

\newpage
\begin{abstract}

We test recent claims that causal (driver/response) relationships can be deduced
from interdependencies between simultaneously measured time series. We apply two 
recently proposed interdependence measures which should give similar results as 
cross predictabilities used by previous authors. The systems which we study are 
asymmetrically coupled simple models (Lorenz, Roessler, and Henon models), the 
couplings being such as to lead to generalized synchronization. If the data 
were perfect (noisefree, infinitely long), we should be able to detect, at least 
in some cases, which of the coupled systems is the driver and which the response.
This might no longer be true if the time series has finite length. Instead, estimated 
interdependencies depend strongly on which of the systems has a 
higher effective dimension {\it at the typical neighborhood sizes} used to 
estimate them, and causal relationships are more difficult to detect.
We also show 
that slightly different variants of the interdependence measure can have quite 
different sensitivities.

\end{abstract}

\end{titlepage}

\newpage
\section{Introduction}

The study of synchronization between chaotic systems has been a topic
of increasing interest since the beginnings of the '90s. 
One important step in this direction was the introduction of the
concept of {\it generalized synchronization} \cite{ott,pikovsky,rulkov}, 
extending previous studies of coupled identical systems ({\it identical
synchronization} \cite{fuji-yamada,fujisaka,piko1,schuster,pecora,pg}) to the 
study of coupled systems with different dynamics.
 
Let us denote by  $\bX$ and $\bY$ two dynamical systems, and by
$\bx=\{x_1,\ldots,x_d\}$ and $\by=\{y_1,\ldots,y_r\}$ their state
vectors, obtained for example by delay embedding.
We assume in the following that the dynamics is deterministic with
continuous time (the case of maps is completely analogous, and will be treated in 
Secs. 3 and 4). 
We further assume the systems are unidirectionally coupled, say  $\bX$
is the autonomous driver and  $\bY$ the driven response
\bea
\dot{\bx}(t) & = & F (\bx(t))\;,            \nonumber \\
\dot{\by}(t) & = & G (\by(t),\bx(t)) 
\label{eq:systems}
\eea

We speak about generalized synchronization between $\bX$ and $\bY$ if the 
following relation exists:
\begin{equation}
\label{eq:function}
\by(t) = \Psi( \bx(t) )
\end{equation}
\noindent
This requirement is less restrictive than the one of
identical synchronization, in which $\Psi = 1$.
Equation (\ref{eq:function}) implies that the state
of the response system is a function {\bf only} of the state of the
driver.  It is not to be confused with the opposite relation 
$\bx(t) = \Phi(\by(t))$ (considered in \cite{quyen2}), which is 
generically valid for sufficiently high
embedding dimensions if the coupling is non-singular, in the sense of
obeying  $\det(\partial G_i/\partial x_n) \neq 0$ everywhere. 
This follows from the implicit function theorem, which allows us to
invert eq.(\ref{eq:systems}b) to $\bx = \chi (\by, \dot{\by})$, and
from the fact that $\tilde{\bf y}(t) = ({\bf y}(t), \dot{\bf y}(t))$ is as good a
state vector as ${\bf y(t)}$.
In particular, if we consider $G (\by(t),\bx(t)) = H(\by(t)) +
U(\bx(t))$, we immediately have 
 $\bx(t) = U^{-1} (\dot{\by}(t) - H(\by(t)))$ if $U(\bx)$ can be inverted. 

The transformation $\Psi$ does not need to be smooth as considered in
\cite{rulkov,abarbanel} and explicitly required in \cite{brown,josic}. 
In fact, Pyragas \cite{pyragas} defined as strong
and week synchronizations the cases of smooth and non-smooth
transformations, respectively (see also \cite{hunt}).

If one of the systems drives the other and a relationship like
equation (\ref{eq:function}) exists, it is possible to predict the
response from the simultaneous state of the driver. But the opposite 
is not true. Just knowing that a relationship like equation
(\ref{eq:function}) exists and that the state of $\bY$ can be predicted 
from that of $\bX$, it is in general not possible to establish 
which is the driver and which is the response.
This is obvious when $\Psi$ is bijective (i.e., $\Phi = \Psi^{-1}$
exists and is unique). The above arguments tell us that $\Psi$ is 
indeed likely to be bijective in case of generalized synchronization,
at least for nearly all $\bx$: If a coupling is not regular in the above sense, 
then its singularities are typically located on a set of measure zero. 
One might tend to believe that $\bX$ must control $\bY$ (and not the opposite), 
if $\by$ follows the motion of $\bx$ with a positive time delay. 
But even then one cannot be sure since there could be an internal delay 
loop in $\bY$ which causes the emitted time series to lag behind. Also, 
both systems could be driven by a third system.
Thus, detecting causal relationships is not easy in general, although it is of 
course of utmost importance in many applications.

In the above we pretended that we could detect {\it exactly} whether 
the state of one of the systems is a function of the other. This is of 
course never the case in practical applications. Different observables 
which should enable one to detect interdependencies in realistic cases
were introduced by several authors. Following an original idea of 
Rulkov et al. \cite{rulkov,abarbanel}, mutual cross-predictabilities were 
defined and studied by Schiff {\it et al.}~\cite{schiff} and
Le van Quyen {\it et al.}~\cite{quyen1,quyen2}. A quantity more closely 
related to that of \cite{rulkov}, but optimized for robustness to noise and 
imperfections in the data was used in \cite{arnhold}. In the latter paper 
also a number of other variants were discussed. Some of these variants were
tested and found to be inferior, but no systematic tests were made.

In contrast to our above discussion, the authors of \cite{schiff} and
\cite{quyen1,quyen2} claimed that driver/response relationships {\it can}
be deduced from such interdependencies. But their proposals, backed by 
numerical studies of simple model systems, were mutually contradictory.
While it was argued in \cite{schiff} that the driver state ${\bf x}$ should be more 
dependent on the response state ${\bf y}$ (i.e., there exists a stronger 
functional dependency $\by\stackrel{\Phi}{\to} \bx$) than vice versa (which is,
as we said, a bit counter intuitive), exactly the opposite was claimed in 
\cite{quyen1,quyen2}. Finally, in \cite{arnhold} it was claimed that 
neither can be expected to be correct in realistic situations with finite 
noisy data, and that it is in general the system with more 
excited degrees of freedom (the more `active' system) that is more 
independent, while the state of the more `passive' system (with less excited 
degrees of freedom) depends on it.

It is the purpose of the present work to settle this question by 
carefully studying simple toy models, including Lorenz, Roessler, and 
Henon systems, using two of the interdependence measures proposed 
in \cite{arnhold}. Basic notions involved in generalized synchronization are 
reviewed in Sec.2. In Sec.3, we recall the operational definition of 
interdependence used in \cite{arnhold}. Numerical results are presented 
in Sec.4, and our conclusion is drawn in Sec.5.

\section{Generalized Synchronization with Exactly Known Dynamics}

While identical synchronization is easily visualized by plotting the difference 
between one of the coordinates of the driver and the corresponding 
coordinate of the response, no similarly simple way exists to detect 
generalized synchronization. Constructing the function $\Psi$ explicitly 
\cite{brown2} might be possible in particularly simple cases, but since 
this will be never exact, it will be never clear whether deviations from 
eq.(2) are due to lack of synchronization or inexactness of $\Psi$.
Instead, the methods of choice in cases where the exact equations of motion 
are known and where arbitrary initial states can be prepared are the 
study of Lyapunov exponents and the identical synchronization of two 
identical response systems differing in their initial conditions.

For the driver/response systems as in eq.(1), one has $d+r$ different 
Lyapunov exponents. Of these, $d$ exponents coincide with those of the 
(autonomous) driver denoted by $\lambda^{(\bX)}_i,\; i=1\ldots d$. The 
other $r$ exponents coincide with those of the response, 
considered as a non-autonomous system driven by the external signal $\bx(t)$
(called {\it conditional Lyapunov exponents} in \cite{pecora}) \footnote{
To see this, we have to recall that all Lyapunov exponents are obtained by 
iterating $d+r$ basis vectors in tangent space, and re-orthogonalize them 
repeatedly. Tangent vectors corresponding to $\lambda^{(\bX)}_i$ span only 
the first $d$ coordinates. The remaining tangent vectors have the first $d$ 
components equal to zero, either by orthogonalization or because their
last $r$ components increase 
faster than any of the first $d$ components.}.
They will be called $\lambda^{(\bY)}_i,\; i=1\ldots r$. Ranking the 
Lyapunov exponents as usual by magnitude, we have generalized synchronization 
iff $\lambda^{(\bY)}_1 < 0$. 

Furthermore, once the Lyapunov exponents are known, the dimension of the 
combined system $\bX+\bY$ can be estimated from the Kaplan-Yorke formula 
\cite{kaplan} 
\be
D_{\bX+\bY} = l + \sum_{j=1}^l \frac{\lambda_j}{|\lambda_{l+1}|}
\label{eq:kaplanyorke}
\ee
(here, $l$ is the largest integer for which the sum over $j$ is non negative).
Generically, we must expect this also to be the dimension of $\bY$ alone
\footnote{I.e., the dimension of an attractor constructed exclusively from 
components of $\by$; we keep of course the fact that $\bY$ is driven by $\bX$.}.
The reason is that, as pointed out in the introduction, $\bX$ will be 
a (single- or multivalued) function of $\bY$, if the inverse of $G(\bx,\by)$
is single- or multivalued. On the other hand, the Kaplan-Yorke dimension 
of $\bX$ alone may be equal to $D_{\bX+\bY}$ or smaller. It is given by a 
formula similar to eq.(\ref{eq:kaplanyorke}) but with $\lambda_i$ replaced 
by $\lambda^{(\bX)}_i$. We see that 
\be
  D_{\bX} < D_{\bX+\bY} \qquad {\rm iff}\qquad \lambda_1^{(\bY)} > \lambda_{l+1}^{(\bX)},
\ee
where $l$ is determined by $\sum_{j\leq l}\lambda_j^{(\bX)} \leq 0 < 
\sum_{j\leq l+1}\lambda_j^{(\bX)}$.
If this inequality holds (together with $\lambda_1^{(\bY)} <0$), we have 
{\it weak synchronization} in the sense of Pyragas \cite{pyragas}. In the 
opposite case, i.e. $\lambda_1^{(\bY)} \leq \lambda_{l+1}^{(\bX)}$, 
one is likely to have strong synchronization, although this 
might not be true due to multifractality: Due to the latter, it is possible 
that the box-counting dimension of $\bX$ is strictly smaller than that of $\bX+\bY$, 
although the equality holds for the Kaplan-Yorke (i.e., information) 
dimensions. In such a case $\Psi$ cannot be smooth, but regions where $\Psi$ 
is non-smooth might well be of measure zero \cite{hunt}.

Another approach for detecting generalized synchronization is by using
two identical response systems which differ only in their initial
conditions. 
If these replicas get
synchronized after some transient, their  trajectories are obviously
independent of the initial conditions, thus being only a function of
the driver. This is most easily checked visually, e.g. by plotting 
the difference between two analogous components of the two replicas 
against time. In this way one can also check for intermittencies and long 
transients which can, together with finite numerical resolution, 
severely obscure the interpretation.

\section{Generalized Synchronization in Real Life}

The above considerations depend on the availability of the exact equations 
of motion, and on the ability to prepare identical replicas. Neither holds
for typical applications.

Real signals usually consist of short segments of data contaminated by noise. 
Furthermore, the dynamics of the system is not known, and therefore the 
methods described in the previous section are not applicable.
  
While identically synchronized systems describe the same 
trajectory in the phase space, the hallmark of a relationship as in eq.(2)
is that any recurrence of $\bX$ implies a recurrence of $\bY$. 
If $\bX$ comes {\it exactly} back to a state it had already been in 
before, the same must be true for $\bY$. In real data, one cannot 
of course expect exact recurrence. We will therefore use as a 
criterion that whenever two states of $\bX$ are similar, the contemporary 
states of $\bY$ are also similar. 

In \cite{schiff,quyen1,quyen2}, this was implemented by making forecasts 
of $\bx_n$ using local neighborhoods (e.g., by means of locally linear maps), and 
comparing the quality of these forecasts with that of forecasts based 
on ``wrong" neighborhoods. In the latter, the nearest neighbors of $\bx_n$ 
are replaced by the equal time partners of the nearest neighbors of $\by_n$.
For reasons explained in \cite{arnhold}, we prefer to use instead a 
measure closer to the original proposal of Rulkov {\it et al.} \cite{rulkov}.
But we should stress that we see no reason why our results should not be 
carried over to the observables used in \cite{schiff,quyen1,quyen2} 
immediately.

Let us suppose we have two simultaneously measured univariate time series
from which we can reconstruct $m$-dimensional delay vectors \cite{takens}
$\bx_n=(x_n,\ldots,x_{n-m+1})$ and
$\by_n=(y_n,\ldots,y_{n-m+1})$, $n=1,\ldots N$.

Let $r_{n,j}$ and $s_{n,j}$, $j=1,\ldots,k$, denote the time indices of the $k$ 
nearest neighbors of $\bx_n$ and $\by_n$, respectively. 
For each $\bx_n$, the squared mean Euclidean distance to its $k$ 
neighbors is defined as

\be
   R_n^{(k)}(\bX)=\frac{1}{k}\sum_{j=1}^{k}{\left( \bx_n - \bx_{r_{n,j}} \right)^2} 
\label{R1}
\ee
and the {\it {\bY}-conditioned} squared mean Euclidean distance is defined by 
replacing the nearest neighbors by the equal time partners of the closest 
neighbors of $\by_n$,

\begin{equation}
   R_n^{(k)}(\bX|\bY)=\frac{1}{k} \sum_{j=1}^{k}{\left( \bx_n - 
          \bx_{s_{n,j}} \right)^2}.                                  
\label{R2}
\end{equation}

If the point cloud $\{\bx_n\}$ has average squared radius $R(\bX) 
= \langle R^{(N-1)}(\bX)\rangle$ and effective dimension $D$ (for a 
stochastic time series embedded in $m$ dimensions, $D=m$), 
then $R_n^{(k)}(\bX) \ll R(\bX)$ for $k \ll N$. More precisely, 
we expect 
\be
    R_n^{(k)}(\bX) / R(\bX) \sim (k/N)^{2/D},  \label{knd}
\ee
where $D$ is the 
dimension of the probability measure from which the points $\bx_n$ are drawn.
Furthermore, $ R_n^{(k)}(\bX|\bY) \approx R_n^{(k)}(\bX) \ll R(\bX)$ 
if the systems are strongly correlated, while 
$R_n^{(k)}(\bX|\bY) \gg R_n^{(k)}(\bX)$ if they are independent.
Accordingly, we can define an interdependence measure
$S^{(k)}(\bX|\bY)$ as 
\be
   S^{(k)}(\bX | \bY) = \frac{1}{N} 
           \sum_{n=1}^N \frac{R_n^{(k)}(\bX)}{R_n^{(k)}(\bX|\bY)} .
\label{SXY}
\ee
Since $R_n^{(k)}(\bX|\bY)\ge R_n^{(k)}(\bX)$ by construction, we have 
\be
   0 < S^{(k)}(\bX | \bY) \le 1.
\ee
Low values of $S^{(k)}(\bX | \bY)$ indicate independence between $\bX$
and $\bY$, while high values indicate 
synchronization (becoming maximal when $S^{(k)}(\bX|\bY)\to 1$).

The opposite interdependence $S^{(k)}(\bY|\bX)$ is 
defined in complete analogy. It is in general {\it not} equal to 
$S^{(k)}(\bX|\bY)$. 
If $S^{(k)}(\bX|\bY) > S^{(k)}(\bY|\bX)$, i.e. if $\bX$ depends 
more on $\bY$ than vice versa, instead of assuming a causal
relationship, we just say that $\bY$ is more ``active" 
than $\bX$. As was argued in \cite{arnhold}, high activity is 
mainly due to a large effective dimension $D$, on the typical 
length scale set by the distances $|\bx_n - \bx_{r_{n,k}}|$ and 
$|\bx_n - \bx_{s_{n,k}}|$.\\

The second interdependence measure to be used in this work was also 
introduced in \cite{arnhold}.
In eq.(\ref{SXY}) we essentially compare the $\bY$-conditioned mean squared 
distances to the unconditioned m.s. nearest neighbor distances. Instead of this, we 
could have compared the former to the m.s. distances to {\it random} points, 
$R_n(\bX) = (N-1)^{-1} \sum_{j\neq n} (\bx_n-\bx_j)^2$. Also, in ergodic theory 
often geometric averages are more robust and more easy to interpret than 
arithmetic ones. Therefore, let us use the geometrical 
average in the analogon of eq.(\ref{SXY}), and define 
\be
      H^{(k)}(\bX | \bY) = \frac{1}{N} \sum_{n=1}^N \log \frac{R_n(\bX)}{R_n^{(k)}(\bX|\bY)}
                                                    \label{HXY}
\ee
This is zero if $\bX$ and $\bY$ are completely independent, while it is positive if
nearness in $\bY$ implies also nearness in $\bX$ for equal time partners.
It would be negative if close pairs in $\bY$ would 
correspond mainly to distant pairs in $\bX$. This is
very unlikely but not impossible. Therefore, $H^{(k)}(\bX | \bY)=0$ suggests that
$\bX$ and $\bY$ are independent, but does not prove it. This (and the asymmetry under 
the exchange $\bX\leftrightarrow \bY$) is the main difference
between $H^{(k)}(\bX | \bY)$ and mutual information. The latter is strictly positive
whenever $\bX$ and $\bY$ are not completely independent. As a consequence,
mutual information is quadratic in the correlation $P(\bX,\bY)-P(\bX)P(\bY)$ for weak
correlations ($P$ are here probability distributions), while $H^{(k)}(\bX | \bY)$ is
linear. Thus $H^{(k)}(\bX | \bY)$ is more sensitive to weak dependencies which 
might make it useful in applications. Also, it should be easier to estimate 
than mutual informations which are notoriously hard to estimate reliably.

\section{Numerical Examples}

The aim of this section is to see numerically whether there exists 
any relationship between the `activity' defined in the last section, 
and a driver/response relationship. In principle there should exist 
such a relationship, since we have argued that the system with higher 
dimension should be more active, and usually the response does have 
higher dimension. This would agree with the conclusion of \cite{schiff}, 
and contradict \cite{quyen1,quyen2}. But it is well known that 
observed attractor dimensions can be quite different from real ones, 
in particular if one has only a finite amount of noisy data and weakly 
coupled systems \cite{lorenz91}.

In order to obtain results which can be easily compared to those of
\cite{schiff,quyen1,quyen2}, we study the same systems as these 
authors.

\subsection{Lorenz Driven by R\"{o}ssler}

As a first example, we studied the unidirectionally coupled systems
proposed in reference \cite{quyen2}. The driver is an
autonomous  R\"{o}ssler system with equations:
\bea
\dot{x}_1 &= & - \alpha \{ x_2 + x_3 \}           \nonumber  \\
\dot{x}_2 &= & \alpha \{x_1 + 0.2 \; x_2 \}      \\
\dot{x}_3 &= & \alpha \{0.2 + x_3 ( x_1 - 5.7 ) \} \nonumber
\label{eq:rossler}
\eea
\noindent
which drives a Lorenz system in which the equation for $\dot{y}_2$ is 
augmented by a driving term involving $x_2$,
\bea
\dot{y}_1 &= & 10 (-y_1 + y_2)                    \nonumber \\
\dot{y}_2 &= & 28 \; y_1 - y_2 - y_1 \; y_3 + C \; x_2^2   \\
\dot{y}_3 &= & y_1 \; y_2 - \frac{8}{3} y_3 .     \nonumber
\label{eq:Lorenz}
\eea

The unidirectional coupling is introduced in the last term of the 
second equation, the constant $C$ being its strength. Notice that 
$x_2$ enters quadratically in the coupling, whence it cannot be 
written as a univalent function of $y_1, y_2, y_3$, and $\dot{y}_2$.
Thus we do not have a strict argument telling us that $D_{\bY}\geq
D_{\bX}$, but the latter seems extremely likely.

As in reference \cite{quyen2} the parameter $\alpha$ is introduced in
order to control the relative frequencies between the two systems. 
The differential equations were iterated, together with the equations 
for the tangent vectors, by using fourth and 
fifth order Runge-Kutta algorithms with $\Delta t = 0.003 - 0.009$. This was
checked to yield numerically stable results, while larger $\Delta t$ and/or
a third order algorithm would have given different results. In 
order to eliminate transients, the first $10^6$ iterations 
were discarded. From the increase of the tangent vectors during 
the following $10^6$ iterations we obtained the Lyapunov exponents. 
Delay vectors with delay $\tau=0.3$ and embedding dimensions 4 and 5 
were constructed from $x_1(t)$ and $y_1(t)$. This delay corresponds 
to roughly 1/4 of the average period of the Lorenz equations. All 
time sequences had length $N=5000$. In order to check for stability 
and for very long transients, all computations were repeated several 
times with different initial conditions.

\begin{figure}
\centering
\psfig{file=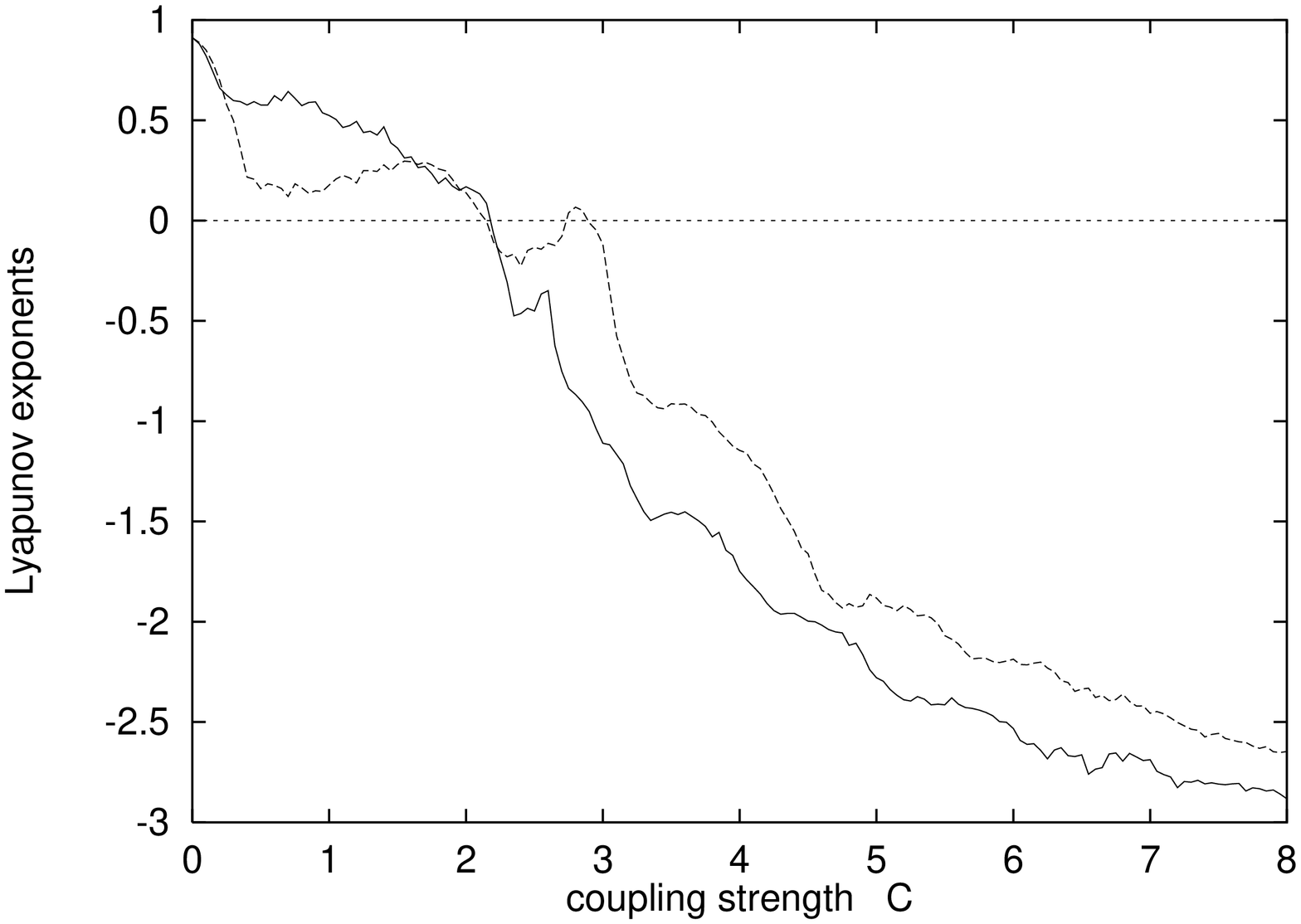,height=10cm,width=6cm,angle=-90}
\caption{Maximum Lyapunov exponents of the modified Lorenz system driven by a R\"{o}ssler. 
The continuous curve is for $\alpha=6$, the broken for $\alpha=10$.}
\label{fig:rololyap}
\end{figure}

For the parameters considered here, the Lyapunov exponents of the 
R\"ossler are $\lambda_i \approx 0.09\alpha,\; 0,\; -5\alpha$; and the
ones of the Lorenz without coupling are $\lambda_i \approx 0.84,\; 0,\; -14.5$.
Figure~\ref{fig:rololyap} shows the maximum Lyapunov exponent of the
driven Lorenz system, as a function of the coupling strength $C$.
The continuous curve shows the result for $\alpha=6$, the 
broken one corresponds to $\alpha=10$. For generalized synchronization,
the maximum Lyapunov exponent should be negative. When $\alpha = 10$, 
this is observed for $2.1<C<2.7$ and for $c>2.9$. 
For all values of $C$ considered in fig.~\ref{fig:rololyap}, the 
maximal Lyapunov exponent of the driven system is larger than the 
smallest one of the driver, and therefore we have only weak 
synchronization.

\begin{figure}
\begin{center}
\psfig{file=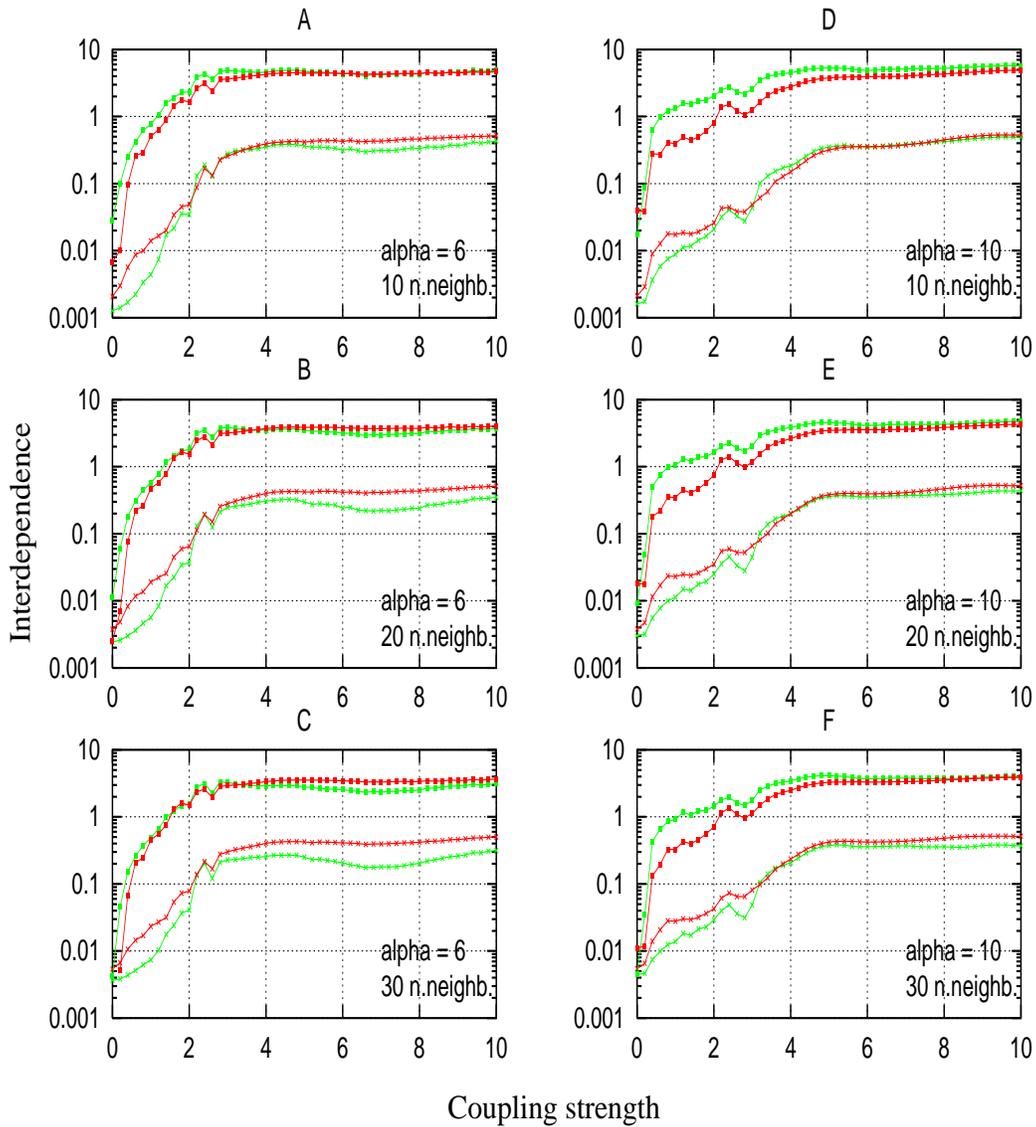,height=15cm,width=14cm,angle=0}
\vspace{0.5cm}
\end{center} 
\caption{Non-linear interdependences between a Lorenz system driven by a
  R\"{o}ssler system, plotted as a function of the coupling strength. 
  Left side graphs correspond to  $\alpha = 6$ and
  right side graphs to  $\alpha = 10$. From top to bottom, the graphs are
  for 10, 20, and 30 nearest neighbors. 
  $S(\bY|\bX)$: black lines with crosses,  $S(\bX|\bY)$: grey lines with
  crosses,  $H(\bY|\bX)$: black lines with squares and  $H(\bX|\bY)$: grey
  lines with squares. See text for details.} 
\label{fig:rolopred}
\end{figure}

Figure~\ref{fig:rolopred} shows the interdependencies $S(\bX|\bY)$,
$S(\bY|\bX)$, $H(\bX|\bY)$ and $H(\bY|\bX)$ for $\alpha = 6$ and $\alpha=10$, and for
different number of nearest neighbors. The embedding dimension was $m=5$.
In all panels, lines with crosses (lower curves) are for $S$ and lines with 
squares (upper curves) are
for $H$; the dark lines denote $(\bY|\bX)$ interdependencies and the grey lines
are for $(\bX|\bY)$. All interdependencies rise with the coupling strength, 
with only few exceptions. These exceptions (at $C\approx 2.5$ for $\alpha = 6$ 
and at $2.3<C<2.8$ for $\alpha = 10$) occur exactly at places where the 
maximal Lyapunov exponent of the driven system is non-monotonic. Thus the 
dependencies are monotonic functions of the Lyapunov exponent. The measure 
$S$ is more sensitive to the sign of the Lyapunov exponent than is $H$, as 
seen from the sharper increase of $S$ when the Lyapunov exponent passes through 
zero and the systems synchronize.

In reference \cite{quyen2} this system was studied only for $\alpha=6$ and 
$C=8$. The latter corresponds to very strong coupling. It was found that 
the interdependence of $\bY$ from $\bX$ was larger than vice versa. This was 
taken as a proof that in general the response depends more on the driver 
than vice versa, and it was proposed that this result could
be used as a general method to detect driver/response relations. 

Our results with $S$ and $H$ agree perfectly with those of \cite{quyen2}, 
if we keep the same values for $\alpha$ and $C$, and use $k\ge 20$, i.e. for 
large neighborhoods. In case of $S$ we find $S(\bY|\bX)\ge S(\bX|\bY)$ for all $k$ 
and $\alpha$, provided $C>5$. Finally, we also find the same 
inequality $S(\bY|\bX)>S(\bX|\bY)$ for very small couplings (below the synchronization 
threshold), while the opposite inequality sometimes holds for intermediate $C$. 

A more consistent picture is seen in the behavior of $H(\bX|\bY)$ and $H(\bY|\bX)$. 
Except in the case $\alpha=6, C>4$, and $k\ge 20$, we always found 
$H(\bX|\bY) > H(\bY|\bX)$, in agreement with the prediction of \cite{schiff}. 
This inequality is most pronounced for small values of $C$.

Our results can be understood by the following heuristic arguments:
\begin{itemize}
\item For strong couplings the two systems are so strongly synchronized 
that the differences in interdependence are small, and we can predict $\bX$ 
from $\bY$ essentially as well as $\bY$ from $\bX$.
\item The theoretical predictions $H(\bX|\bY) > H(\bY|\bX)$ and $S(\bX|\bY) > S(\bY|\bX)$
are based on the limiting behavior for small neighborhoods. It is thus 
not too surprising that they can be violated for large values of $k$. 
\item For uncoupled systems ($C=0$) one has $S(\bX|\bY) < S(\bY|\bX)$ if 
$D_{\bf X} < D_{\bY}$ and vice versa. Notice that this (which is easily obtained 
from eq.(\ref{knd})) is the opposite of what we expect if 
$D_{\bf X} < D_{\bY}$ holds only due to the coupling. 
In our case, $D_{\rm Roessler}  \approx 2.018 < D_{\rm Lorenz}  \approx 2.058$.
This explains why $S(\bX|\bY) < S(\bY|\bX)$ for very small $C$ ($C<1$). 
In contrast, $H(\bX|\bY) =H(\bY|\bX)=0$ for uncoupled systems, whence no such 
problem exists for $H$. Thus we expect that the behavior of $H(\bX|\bY)$ at small 
couplings is easier to interpret than the behavior of $S$ which should depend
nontrivially on $k$ and $N$. This is precisely what we found. 

\end{itemize}

\subsection{Two Coupled Henon Maps}

As a second example we studied two unidirectionally coupled Henon
maps similar to the ones proposed in \cite{schiff}, with equations

\bea
x'_1 &=& 1.4 - x_1^2 + b_1 \; x_2          \nonumber \\
x'_2 &=& x_1    
\label{eq:henon1}
\eea

\noindent
for the driver, and

\bea
y'_1 &=& 1.4 - (C\;x_1y_1 + (1-C)\;y_1^2) + b_2 \; y_2          \nonumber \\
y'_2 &=& y_1    
\label{eq:henon2}
\eea

\noindent
for the response. Again we discarded the first $10^6$ iterations, and obtained 
Lyapunov exponents from the next $10^6$. Interdependencies were 
then estimated from $N=5000$ iterations, using 3-dimensional delay 
vectors.
As in the previous example, the stability of the results was checked 
by starting from different initial conditions.
For calculating the interdependencies 
we used $k=10$ nearest neighbors. No significant changes were observed 
for other values of $k$.

The constants $b_1$ and $b_2$ were both set
to 0.3 when analyzing identical systems, and to 0.3, 0.1 when
analyzing non-identical ones. Furthermore, in all cases we also studied 
how the results changed if white measurement noise was added either to 
the driver, to the response, or to both.

\subsubsection{Identical systems}

We first studied the case $b_1=b_2=0.3$. This is the ``canonical"
value for the Henon map, for which $\lambda_1=0.4192$ and $D=1.26$. 
One easily sees that $(y_1,y_2)=(x_1,x_2)$ 
is a solution of eqs.(\ref{eq:henon1},\ref{eq:henon2}). Thus we 
can have identical synchronization \cite{pecora}, but due to the 
asymmetry of the coupling we cannot rule out non-identical (generalized) 
synchronization either.

Figure~\ref{fig:hehelyap} (solid line) shows the maximum Lyapunov 
exponent of the response system.
It becomes negative for couplings larger than 0.7, when identical
synchronization between the systems takes place. But it is also 
slightly negative for $0.47<C<0.52$. Plotting differences $x_1-y_1$, 
e.g., one sees that there is no identical synchronization in this 
window. But making a cut through the attractor of the combined 
system, by plotting e.g. pairs $(x_2,y_2)$ whenever $|x_1|<10^{-4}$, 
one sees a fractal structure which clearly shows that there is no 
identical synchronization. On the other hand, $\lambda_1^{(\bY)}<0$ which 
leaves only the possibility of generalized 
synchronization. In this window $(0.47<C<0.52)$, the Kaplan-Yorke 
formula gives $D\approx 2.25$, showing that this synchronization is weak. 
For $C>0.7$, the Kaplan-Yorke formula 
does of course not apply to the combined system and $D=D_{\rm Henon}=1.26$.

Interdependencies are shown in fig.~\ref{fig:hehepred}A and fig.~\ref{fig:hehepred}D. 
As expected, we see that $S$ and $H$ both rise sharply at $C\geq 0.7$, where 
identical synchronization sets in. The fact that synchronization is perfect 
is seen from the fact that $S(\bY|\bX)=1$ for $C\geq 0.7$. In contrast, we 
do not see any anomaly for $C\approx 0.5$, showing again that the synchronization 
at $c\approx 0.5$ is very weak indeed. 
For $C<0.7$ we see that $H(\bX|\bY) > H(\bY|\bX)$, in agreement
with our general prejudice that the response has higher dimension and is 
thus more active. Although not so pronounced, this difference is also
seen for $S$.

We also analyzed how both measures changed with the inclusion of
measurement noise ($S/N$ amplitude ratio $ \sim 25 \% $).
Figs.~\ref{fig:hehepred}B and 4E show the results when the noise
is added to the driver and figs.\ref{fig:hehepred}C and 4F when it is 
added to the response. In general, we expect of course a decrease 
of any dependences when noise is added. Formally, we have to discuss 
how $R^{(k)}(\bX|\bY)$, $R^{(k)}(\bY|\bX)$, $R^{(k)}(\bX)$, and 
$R(\bX)$ change if noise is added to $\bX$. 
\begin{itemize}
\item $R(\bX)$ changes very little, since $S/N \ll 1$;
\item $R^{(k)}(\bX)$ increases strongly, since the dimension of the 
noisy time series is large;
\item $R^{(k)}(\bX|\bY)$ increases little if $\bX$ and $\bY$ are weakly 
dependent (it is already large), but increases even more than $R^{(k)}(\bX)$ 
if $\bX$ and $\bY$ are strongly dependent. 
\item The last is also true for $R^{(k)}(\bY|\bX)$.
\end{itemize}

From these we see that $H(\bY|\bX)$ should decrease roughly as much as 
$H(\bX|\bY)$, and this decrease should be strongest if $\bX$ and $\bY$ 
are fully synchronized. In contrast, $S(\bX|\bY)$ should decrease much less 
(or even increase) if noise is added to $\bX$, and if $\bX$ and $\bY$ are 
weakly dependent. If $\bX$ and $\bY$ are fully synchronized, adding noise 
to $\bX$ suppresses $S(\bY|\bX)$ much more than $S(\bX|\bY)$ since the 
strong increase of $R^{(k)}(\bY|\bX)$ is then not compensated by any change 
of $R(\bY)$. Adding noise to $\bY$ can be discussed similarly.
All these predictions are fully verified in fig.4. Notice that measurement 
noise can reverse the general inequality $H({\rm driver}|{\rm response})
> H({\rm response}|{\rm driver})$, as seen e.g. in fig.\ref{fig:hehepred}E
for $C>0.7$.

In general, $S$ seems to be less robust against measurement noise than $H$.

\subsubsection{Non identical systems}

\begin{figure}
\begin{center}
\psfig{file=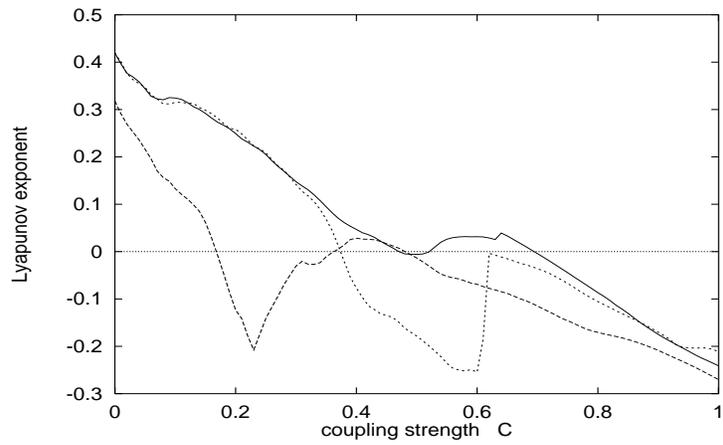,height=10cm,width=6cm,angle=-90}
\end{center}
\caption{Maximum Lyapunov exponent of the response system for two coupled Henon systems 
with $b_1 = b_2=0.3$ (solid line); $b_1 = 0.3, b_2=0.1$ (dashed line);
and $b_1 = 0.1, b_2=0.3$ (dotted line).}
\label{fig:hehelyap}
\end{figure}

\begin{figure}
\centering
\psfig{file=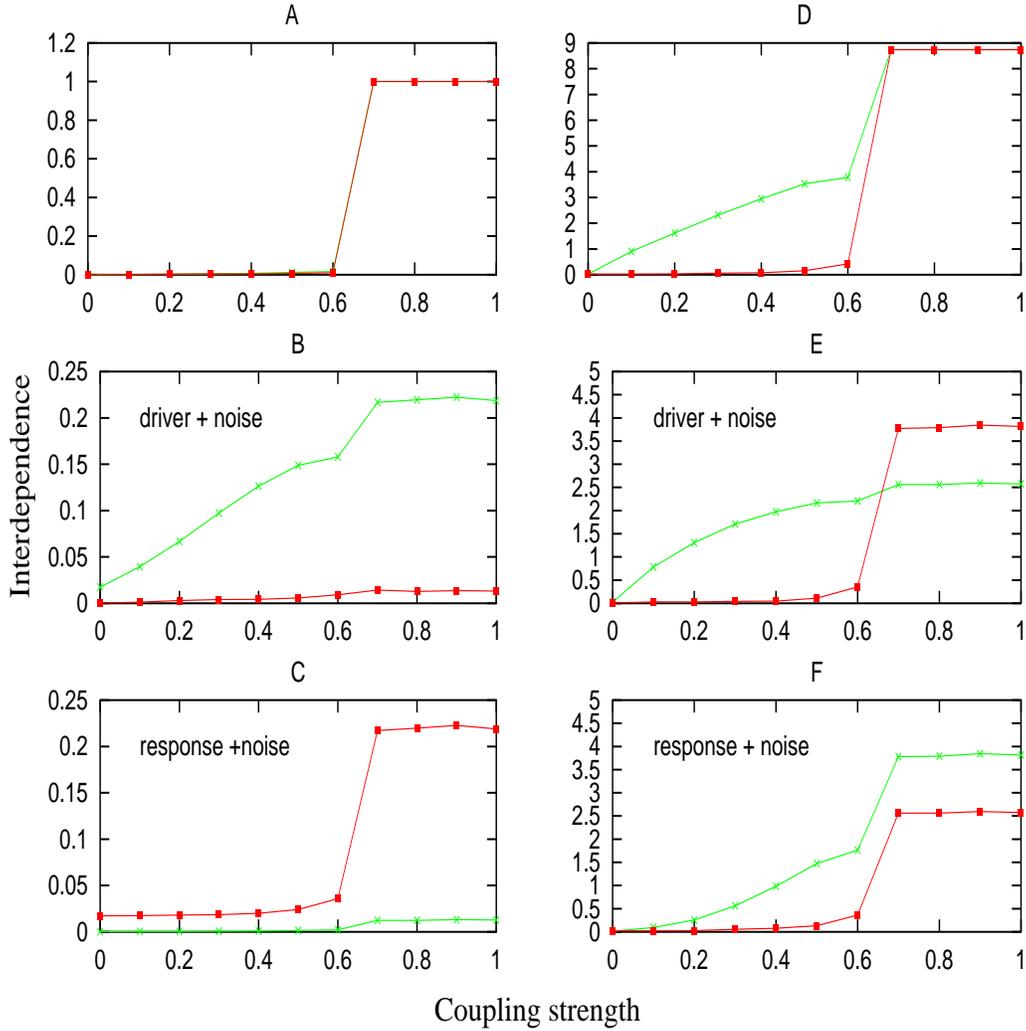,height=14cm,width=14cm,angle=0}
\caption{Interdependencies between two Henon systems with equal $b$-parameter, 
   $b_1 = b_2=0.3$ as functions of the coupling strength 
   (panels A and D). In the other panels, white measurement 
   noise has been added either to the driver (panels B, E) or to the response 
   (panels C, F). Left side figures correspond to $S$ and the ones
   in the right side to $H$. Black lines are the $(\bY|\bX)$
   interdependencies and grey lines correspond to $(\bX|\bY)$.}
\label{fig:hehepred}
\end{figure}

\begin{figure}
\centering
\psfig{file=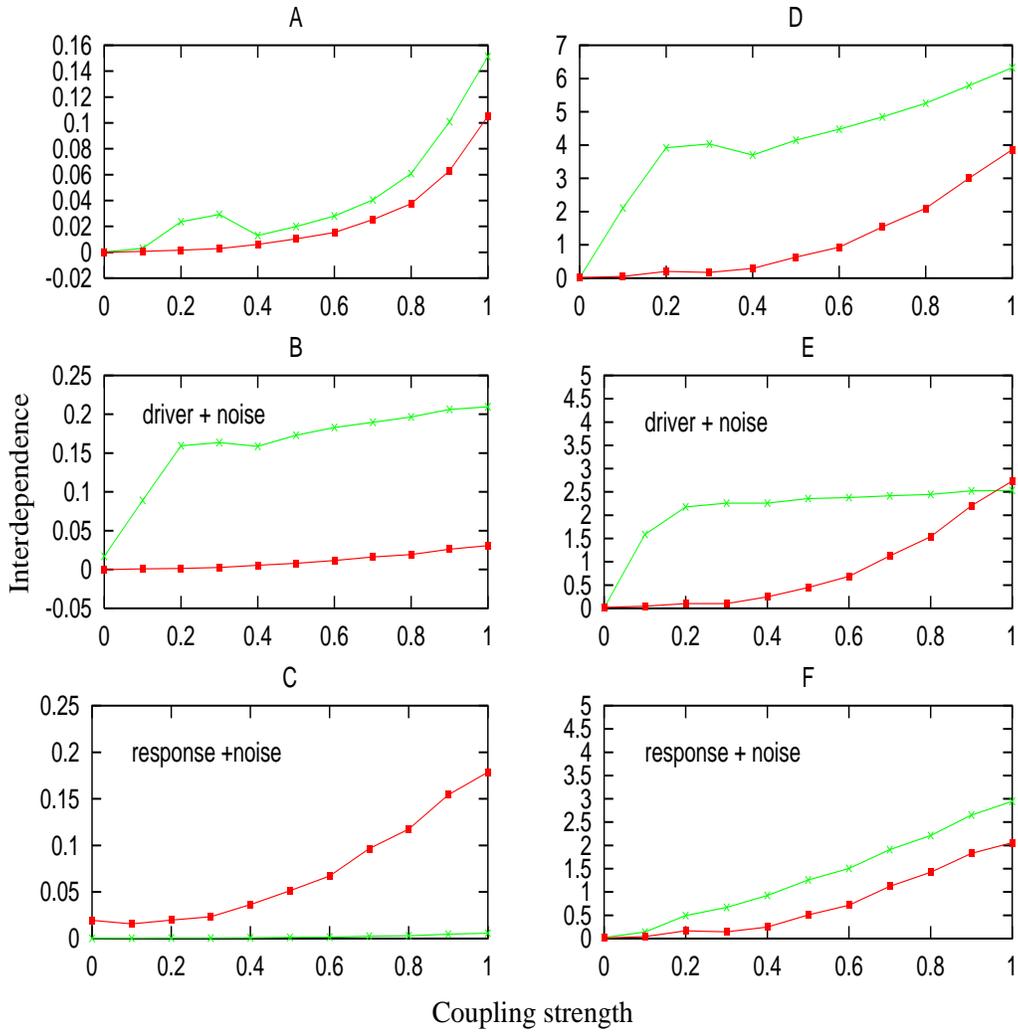,height=14cm,width=14cm,angle=0}
\vspace{0.6cm}
\caption{Same as fig.\ref{fig:hehepred}, with $b_1 = 0.3$, $b_2 = 0.1$.}
\label{fig:hehepred1}
\end{figure}

Figures~\ref{fig:hehepred1}A and 5D show dependencies for different 
$b$-parameters ($b_1=0.3, b_2=0.1$) where identical synchronization 
is impossible, and where the driver has higher dimension than the 
uncoupled response. 
In this case, the interdependencies do not increase as sharply as 
in the previous case, and they reach lower values.
With both measures we see an increase between $C=0.1-0.4$ in agreement
with the negative values of the maximum Lyapunov exponent for these
coupling strengths (see fig.\ref{fig:hehelyap}). 
As in reference \cite{schiff}, we found $S(\bX|\bY) > S(\bY|\bX)$ and
$H(\bX|\bY) > H(\bY|\bX)$.
These inequalities still hold when adding measurement noise to the driver
(fig.\ref{fig:hehepred1}B,E). But when the noise is added to the
response (fig.\ref{fig:hehepred1}C,F), only the inequality for 
$H$ survives, while that for $S$ is reversed.  These dependences on noise can be 
discussed in complete analogy with the previous case of identical systems.

\begin{figure}
\centering
\psfig{file=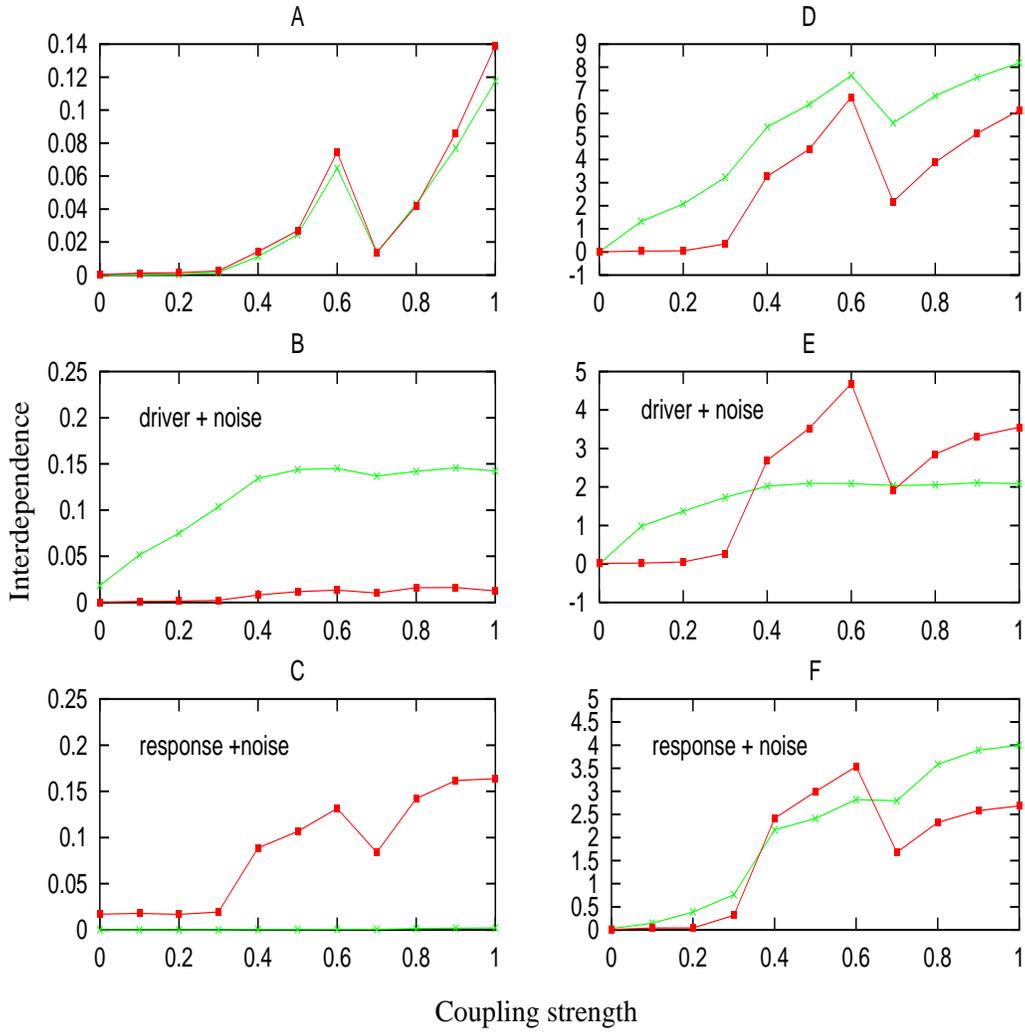,height=14cm,width=14cm,angle=0}
\vspace{0.6cm}
\caption{Same as fig.\ref{fig:hehepred} with $b_1 = 0.1$, $b_2 =
  0.3$. In panel C, the curve for $S(\bX|\bY)$ coincides with the x-axis within our precision.}
\label{fig:hehepred2}
\end{figure}

The situation changes slightly if the uncoupled response has higher 
dimension than the driver, as in the case $b_1=0.1, b_2=0.3$ shown 
in fig.~\ref{fig:hehepred2}. The panels of this figure show more 
structure than those of fig.\ref{fig:hehepred1}, mainly since also the 
Lyapunov exponent shows more structure (see fig.\ref{fig:hehelyap}).
From fig.~\ref{fig:hehepred2}F we see that there is a parameter window, 
$0.4\leq C\leq 0.6$, where $H(\bY|\bX) > H(\bX|\bY)$ after adding 
noise to $\bY$. This is not yet understood, but all other features in this 
plot can be understood heuristically along the lines discussed above.

\section{Conclusion}

In this work we studied the possibility of predicting driver/response 
relationships from asymmetries in nonlinear interdependence measures. 
More precisely, we studied two particular interdependence measures, 
introduced in \cite{arnhold}, and applied them to simple asymmetrically 
coupled strange attractors.
In contrast to previous works, we find that such predictions are not always 
reliable, although we agree with \cite{schiff} that they would be possible 
for ideal (noise-free, infinitely long) data. We agree with 
\cite{quyen1,quyen2} as far as one of their numerical examples is 
concerned, but we show that this was a mere coincidence, and cannot 
be generalized.

Instead, we confirmed the conjecture of \cite{arnhold} that asymmetries 
in interdependence measures reflect mainly the different degrees of 
complexity of the two systems {\it at the level of resolution at 
which these measures are most sensitive}. For practical applications, 
this is not the infinitely fine level at which theoretical arguments 
like those of \cite{schiff} apply. The latter predict correctly 
that the response is more complex, i.e. has a higher Kaplan-Yorke dimension. 
But this argument can become irrelevant even for the extremely simple 
toy models which we studied in the present paper. It should be even 
less relevant in realistic situations where all sorts of noise, 
non-stationarity, and shortness of data present additional 
limitations.

Nevertheless, we propose that asymmetries of measured interdependencies can
be very useful in understanding coupled systems. Indications for this were 
given in \cite{schiff,quyen1,quyen2,arnhold}. All these papers were 
dealing with neurophysiology. Even if no causal relationships can be 
deduced from such asymmetries, it was found in \cite{quyen1,quyen2,arnhold}
that the resulting patterns are closely related to clinical observations, 
and could e.g. contribute to a more precise localization of epileptic
foci and might be useful for predicting epileptic seizures.

An unexpected result of our study is that for the simulations
performed the measure $H(\bX|\bY)$ which 
had not done very well in preliminary tests is actually more robust and 
easier to interpret than the measure $S(\bX|\bY)$ which was mostly
used in \cite{arnhold}.
However, this should not be directly extended to real life data. 
A more systematic comparative study with a large database of EEGs from
 epilepsy patients is under way.

\bibliography{synchro}

\newpage

\newpage

\newpage

\newpage

\newpage

\newpage

\end{document}